\documentstyle[epsf,12pt]{article}
\begin{document}

\begin{center}
{\Large \bf Analysis of the sensor characteristics of the Galileo dust
detector with collimated Jovian dust stream particles} 

\bigskip

{\bf
        H.~Kr\"uger, E. Gr\"un, A. Heck, S. Lammers
}

\bigskip

Max-Planck-Institut f\"ur Kernphysik, Saupfercheckweg 1, 69117 Heidelberg, Germany

\bigskip
\bigskip

\today
\end{center}

\bigskip

\begin{abstract}

The Dust Detector System onboard Galileo records dust impacts 
in the Jupiter system. Impact events are classified into four 
quality classes. Class~3 -- our highest quality class -- has always
been noise-free and, therefore, contains only true dust impacts. 
Depending on the noise environment, class~2 are 
dust impacts or noise. Within $\rm 20\,R_J$ from Jupiter 
(Jupiter radius, $\rm R_J = 71,492\,km$) class~2 shows clear indications 
for contamination by noise. We analyse the dust data 
from Galileo's prime Jupiter mission (1996 and 1997), separate 
dust impacts from noise events and derive 
a complete denoised set of Galileo dust data (class~2 and class~3). 
Collimated streams of nanometer-sized dust particles which have been 
detected throughout the Jovian system (Gr\"un et al. 1998, {\em JGR}, 
103, 20011-20022) are used to analyse the sensitive area and the 
field of view of the dust detector itself. The sensitive area
for stream  particles which trigger class~3 events is 
$\rm 110\,\pm\,37\,cm^2$. This is almost a factor of ten smaller than 
the total sensitive area for class~2 impacts 
($\rm 1,000\,cm^2$). Correspondingly, the field of view of the detector
for class~3 stream particles is reduced from $\rm 140^{\circ}$ to 
$\rm 96^{\circ}$.
The magnetometer boom and other instruments on board 
Galileo cause a significant shadowing of the field of view of
the dust sensor.
Our analysis is supplementary to ground calibrations of the dust 
instrument because the low masses and high speeds of the stream
particles could not be achieved in the laboratory. Our new 
results have important consequences for the analysis of dust 
in the Jupiter system.

\end{abstract}

\section{Introduction}

On 7 December 1995 the Galileo spacecraft arrived at Jupiter and was 
deflected into a highly eccentric orbit about the planet. Galileo
carries a dust impact ionization detector on board (Gr\"un et al. 1992a). 
Dust particles which hit the sensor are recognized by up to three 
independent charge signals (Fig.~\ref{ddsconf}). Impact events are classified 
into four quality classes and six ion charge ($\rm Q_I$) amplitude ranges 
(Gr\"un et al. 1995a, hereafter Paper~I, Kr\"uger et al. 1998a) which lead 
to 24 individual categories. Each impact is counted by one of 24 
accumulators. No indications for contamination by 
noise have ever been seen in class~3, our highest quality class.
In interplanetary space,
class~2 was also noise-free, whereas classes 1 and 0 
contained noise events in their lowest ion amplitude ranges 
(Baguhl et al. 1993, Gr\"un et al. 1995b,c, hereafter Papers~II and III). 

During Galileo's passage through the high radiation environment in 
the inner Jovian magnetosphere on 7 December 1995, however,  class 1 and
class 2 showed indications for noise contamination (Kr\"uger et al. 1998a, 
hereafter Paper~IV): these events occurred at all sensor pointing 
directions. Data obtained later during Galileo's 
orbital tour about Jupiter showed the same behaviour in the class~0, class~1 and 
class~2 events when Galileo was in the inner Jovian system, within 
about $\rm 20\, R_J$  from Jupiter 
(Jupiter radius, $\rm R_J = 71,492\,km$). In contrast to the behaviour of 
events in these lower classes, such a wide spread in impact direction
was not seen for class~3.
The noise events in the lower quality classes 
are probably caused by high energetic electrons in Jupiter's magnetosphere. 

Throughout the Jovian system Galileo has detected collimated streams of tiny
dust particles (Gr\"un et al. 1997, 1998). The behaviour of the impact 
direction of these particles shows significant differences depending upon 
whether they are recognized as class~2 or 
class~3 impacts. No such difference has ever been
observed for particles measured with the Galileo and Ulysses 
dust sensors in interplanetary space. From the instrumental point of 
view the main difference between 
class~2 and class~3 is that class~3 events have three charge signals 
(electron, ion and channeltron signal) whereas only two charge signals 
are required for class~2 (Paper~IV). Most of the class~2 impacts 
have no channeltron signal.

In this paper we investigate the noise in the class~2 data and 
derive a 'denoised' data set which contains class~2 dust impacts 
together with class~3 impacts. Class~3 has always been noise-free. This 
complete data set of impacts from Galileo's orbital tour about Jupiter
is then used for an analysis of the field of view of the dust sensor itself.
 
\section{Noise characterization} \label{noise}

The Galileo spacecraft configuration is shown in Fig.~\ref{galileo}.
Galileo is a dual-spinning spacecraft and the rotation angle (ROT) measures 
the viewing direction of the dust detector (DDS) at the time of a dust particle 
impact. During one spin revolution of the spacecraft the rotation angle scans 
through a complete circle of $360^{\circ}$. At rotation angles of 
$90^{\circ}$ and $270^{\circ}$ the sensor axis lies nearly 
in the ecliptic plane, and at $0^{\circ}$ it is closest to ecliptic north. 
DDS rotation angles are taken positive around the negative spin axis of 
the spacecraft. This is done to easily compare Galileo impact direction
data with those taken by Ulysses which, unlike Galileo, has the opposite 
spin direction.

Figure~\ref{noise1} shows the rotation angle for class~2 and class~3 events 
detected in the lowest ion amplitude range for Galileo's third orbit about 
Jupiter (C3). During this orbit Galileo had a close encounter with Callisto.
During the other 
orbits of Galileo about Jupiter the rotation angle showed a very similar 
behaviour (Gr\"un et al. 1998) and we use C3 as a typical example.
The spacecraft trajectory for the C3 orbit is shown in Gr\"un et al. (1998)
and is not repeated here.

The upper panel in Fig.~\ref{noise1} shows class~3 dust impacts.  When 
Galileo approached the inner Jovian system dust impacts were detected 
at a rotation angle of about $270^{\circ}$ (day 303 to 309). About a day after 
Callisto closest approach (day 309.57) the rotation angle   shifted by 
$180^{\circ}$ and the dust particles were detected with rotation angles
of $90^{\circ}$ 
(day 310 and 311).  A few particles were again detected with rotation 
angles of $270^{\circ}$ on day 314. This behaviour of class~3 dust impacts 
has been discussed in detail by Gr\"un et al. (1997, 1998). It can 
be explained by streams of nanometer-sized dust particles which are ejected from 
the inner Jovian system with very high velocities (about $\rm 300\,km\,s^{-1}$, 
Zook et al. 1996).
Io has recently been identified as the probable  source for these 
particles (Kr\"uger et al. 1998b, A. Graps et al. in prep.). 

In the second panel of Fig.~\ref{noise1} we show the class~2 events 
detected in the same time period. The number of class~2 events shown 
is apparently lower than the number of class~3 impacts. This, however, is 
an effect of the classification scheme used in the 
dust detector. The full information of class~3 events is preferentially 
transmitted to Earth. From the accumulator data we know that
class~2 events are in reality more abundant than class~3 impacts. 

During approach to the inner 
Jovian system the class~2 events were detected with rotation angles of $270^{\circ}$
which is identical with the 
behaviour of the class~3 impacts at this time (days 303 to 
309). Right after the C3 encounter, however, the approach direction of the 
class~2 events 
widened and they were sensed with all rotation angles until day 313. 
This spread in rotation angle occurred in the inner Jovian system when 
Galileo was within about $\rm 20\, R_J$  from Jupiter (days 310 to 313). The 
same occurred during all other passages of Galileo through the inner Jovian 
system, which are not shown here. In this time interval the rotation angle
pattern of class~2 differs
completely from that of class~3. We interpret this marked difference 
between events in both quality classes within $\rm 20 R_J$ from Jupiter as a clear 
indication for 
noise contamination in class~2 in this distance range. This is 
in contrast to the behaviour further away from Jupiter (days 303 to 309) where 
both classes show a very similar rotation angle pattern which 
indicates that here class~2 is very 
likely noise-free and contains mostly true dust impacts.

This marked difference between class~2 and class~3 events 
during impact or noise events (charge amplitudes, charge rise times, time differences 
and coincidences between the charge signals) has been used to derive criteria for a clear 
noise identification. Figure~\ref{noise2} shows two of the parameters investigated: 
1. the difference between the charge measured on the target, $\rm Q_E$, and that 
measured on the ion collector grid, $\rm Q_I$, $\rm EA-IA $ (in digital units), 
and 2. the rise time $\rm t_E$ of the target signal (ET in digital units).

The class~3 impacts show a 
very small spread in $\rm EA-IA$ with a mean value of about 4. This is in 
agreement with laboratory measurements at the Heidelberg dust accelerator which,
for dust impacts, showed target signals usually 2 to 6 units larger than the 
ion charge signals: $\rm 2\leq (EA-IA) \leq 6$.  Only four class~3 impacts have
values larger than ten, and there are only three impacts with a 
value of one or smaller. Outside $\rm 20\, R_J$ (day 303 to 309) the class~2 
events show the same behaviour in $\rm EA-IA$ as class~3 impacts. Within $\rm 20\, R_J$,
however, $\rm EA-IA $ shows a much larger spread in class~2. Very large values
exceeding 20 and small values of one or smaller are frequent. 

The rise time values, ET, of the target signal show a very similar 
characteristic: outside $\rm 20\, R_J$ the digital values 
measured for class~2 cluster 
between 9 and 14 which is the same range as that measured 
for most of the class~3 impacts. In the inner Jovian system, however, all 
possible digital rise time values between 0 and 15 have been found. A very 
similar boundary for class~2 events at $\rm 20\, R_J$ has also been found for 
other parameters which are not shown here:
the rise time of the ion collector signal, $\rm t_I$,
the time difference between the target and the ion collector signals, $\rm t_{EI}$, 
the time difference between the target and the entrance grid signals, $\rm t_{PE}$,
the charge signal measured at the channeltron, $\rm Q_C$, etc.

In order to separate noise events from true dust impacts in class~2 we have 
assumed that class~3 is noise-free and derived boundaries for the 
various impact parameters from the spread of the above-listed parameters for class~3 
far away from Jupiter. All events which fulfill the criteria listed in 
Tab.~\ref{criteria} are classified as noise events. All other events are considered
true class~2 dust impacts. The criteria applied to events in 
the lowest amplitude range AR1 are different from those in the higher amplitude 
ranges AR2 to AR6. The events classified as noise by the criteria listed in 
Tab.~\ref{criteria} are shown in the third panel of Fig.~\ref{noise1}. They 
are strongly concentrated towards the time of perijove passage,
and their rotation angle distribution is nearly isotropic. This is in clear 
contrast to the rotation angle pattern of the class~3 impacts in this time 
interval (days 310 to 313, top panel of Fig.~\ref{noise1}).  

The criteria given in Tab.~\ref{criteria} separate rather well dust 
stream particles from probable noise events. We derived these
criteria from a pure analysis of the data rather than a consideration of 
physical mechanisms that can trigger noise events in the dust sensor. Future
investigations have to show if these criteria can also successfully be 
applied to other populations of dust particles like, for example, the 
secondary ejecta 
particles we have detected close to the Galilean satellites (Gr\"un et al. 
1998).

\section{The 'denoised' Galileo dust data}  \label{denoising}

Class~2 dust impacts after denoising are shown separately in the 
fourth panel of Fig.~\ref{noise1}. They are combined with class~3 impacts 
in the fifth panel. The 'swarm' of class~2 events detected with all rotation 
angles close to perijove passage has nearly completely disappeared. Before
day 309 and on day 314 class~2 impacts show nearly the same behaviour 
as class~3 which indicates that the noise separation derived in the
previous section is quite reliable. The gap 
in class~3 on day 310 is filled with class~2 impacts which have  
a large spread in rotation angle. This is a clear difference between 
class~2 and class~3, and it is a real difference. A similar 
spread in class~2 was also seen during most of the other satellite 
encounters when Galileo was in the inner Jovian system. This feature 
will be discussed in Sect.~\ref{direction}.

2,822 class~2 events detected during Galileo's Jupiter mission in 1996 and 1997
(G1 to E12 orbits) fulfill the noise criteria in Tab.~\ref{criteria}.
Given a total number of 7,507 class~2 events detected by the dust sensor 
in the same time period 
(Tab.~\ref{eventnumber}) the fraction of noise events in class~2 is 
38\%. The complete data set for 1996 and 1997 now contains 10,660 true 
class~2 (4,685) and class~3 (5,975) dust impacts. 97\% of these are small dust
particles detected in the lowest amplitude range AR1.

The noise separation criteria developed in the present analysis have been applied 
to class~1 and class~0 events without success. More than 80\% (5,302) of all events in 
these two classes are in the lowest amplitude range AR1 which was very 
sensitive to noise even in the low-noise interplanetary environment.
Therefore, the identification of true dust impacts in these two 
categories in the data obtained within the Jovian system will be very 
difficult. The remaining 20\% (1,076) of class~0 and class~1 events
in higher amplitude ranges almost completely populate the second amplitude 
range AR2. Even if dust impacts could be identified in these data, this 
would only marginally increase the total number of 10,660 dust impacts 
in class~2 and class~3. We therefore consider class~2 and class~3
impacts  as the complete set of dust data from Galileo's orbital tour
about Jupiter.

A very small number of class~3 events detected in the inner Jovian system 
also fulfills the noise criteria for class~2 listed in Tab.~\ref{criteria}
as can be seen in Fig.~\ref{noise2}. A future analysis has to show if this
is indeed noise. The total number of such class~3 events in the 
data set from the Jovian system is expected to be less than one percent
and, theerfore, is probably irrelevant.

During Galileo's Jupiter mission the memory of DDS is typically read out 
every 7 or 21 minutes and the data are directly transmitted to Earth. 
The latest event detected between two readouts in each of the 
six amplitude ranges plus one older event from any one of the six 
amplitude ranges is transmitted with its full information (charge
signals, rotation angle etc. measured during the impact or noise event, 
Paper~I).
Because the stream particles populate only the lowest amplitude range 
(AR1), some of them are lost if there is more 
than one event between two readouts, or, in other words, if the 
event rate exceeds one per 7 or 21 minutes. This occurs quite 
frequently during Galileo's orbital tour in the Jovian system and only 
a fraction of all particles can be transmitted with their full information.
Therefore, the impact rate has to be determined from the accumulators of the 
instrument (Paper~I) since every impact or noise event is counted with one of 
24 accumulators irrespective of whether it is transmitted to Earth with its 
full information 
or not. The accumulators, however, cannot discriminate between noise and 
true dust impacts. Hence, rates deduced from the class~2
accumulators contain noise especially in the inner Jovian system.

With dust impacts now being separable from noise events, the class~2 event 
rate deduced from the class~2 accumulators (Paper~I) can be 'cleaned' from
noise. The dashed line in Fig.~\ref{noise3} 
shows the event rate (noise plus dust impacts) deduced from the class~2, 
AR1 accumulator. By calculating the ratio between the number of noise events 
and the total number of events (true dust impacts plus noise events)
within a certain time interval, 
an empirical noise correction (relative noise fraction $\rm f_{noi}$) can be 
applied to the impact rate. The corrected impact rate is shown as a 
solid line in Fig.~\ref{noise3}. Each rate value 
has been corrected by the relative noise fraction determined from an interval 
half a day before until half a day after the time of the rate point 
to be corrected. The largest noise correction occurred around 
perijove passage.

\section{The rotation angle pattern of small class~2 and class~3 
dust particles} \label{direction}

In this section we use the denoised dust data set derived in the previous 
section to analyse the field of view (FOV) of the dust sensor for class~2
and class~3 impacts.

\subsection{Spacecraft configuration}
 
To understand the directional behaviour of the impacts one has to 
know the FOV of DDS and the configuration of the 
Galileo spacecraft. The spacecraft configuration is shown in 
Figure~\ref{galileo}. The DDS instrument is mounted underneath the 
magnetometer boom (MAG, Kivelson et al. 1992).
The Plasma Subsystem (PLS, Frank et al. 1992) and 
the Energetic Particles Detector (EPD, Williams et al. 1992) are 
mounted sideways to the boom. The 140$^{\circ}$ FOV of DDS is indicated. 
Note that the magnetometer boom, PLS and EPD are in the FOV of DDS. The
significance of the obscuration by these instruments becomes 
obvious in Fig.~\ref{ddsfov} which shows the FOV of DDS seen by an 
imaginative observer 'sitting` in the center of the target (i.e. on the 
sensor axis). The parts of the FOV obscured by the 
boom, PLS and EPD are at the top. In the following the angle w.r.t. the 
sensor axis will be called impact angle. Note that the boom 
obscures the FOV in -Z direction down to an impact angle of $\rm 33^{\circ}$.

DDS is mounted underneath the magnetometer boom at an angle of 60$^{\circ}$ with
respect to the positive spin axis (+Z direction, JPL Galileo document
GLL-3-180C, page 70). An angle of 55$^{\circ}$ has been
erroneously stated before. Simulations of the Jupiter dust stream
particles show that the instrument viewing geometry which corresponds to
a 60$^{\circ}$ angle gives better agreement with the observed impact 
directions of the particles than that corresponding to a 55$^{\circ}$ 
angle (Heck 1998).

\subsection{Data}

In order to compare the directional behaviour of the class~2 and class~3
dust impacts we use data from the Callisto 9 (C9) orbit as a representive 
example. Similar features have been seen during most of the other orbits 
of Galileo through the inner Jovian system. Figure~\ref{data} shows the 
impact rates and the rotation angle for small class~2 and class~3 impacts 
when Galileo was approaching the inner Jovian system. The trajectory of 
the C9 orbit is shown in Fig.~\ref{c9orbit}. 
The most important features in the impact rates of both
classes which are relevant for this analysis are: 
The rate of class~2 impacts increased by more than an 
order of magnitude on day 174.0. Such a strong increase occurred in the 
class~3 impact rate about 1.4 days later (175.4). Furthermore, 
the impact rate of class~3 shows a gap on day 178.0 to 178.4 which is 
not seen in class~2. A similar gap has also been 
reported for the class~3 rate detected during the first four orbits of 
Galileo's mission about Jupiter (Gr\"un et al. 1998).

The bottom panel of Fig.~\ref{data} displays the rotation angle (ROT) of 
the dust particles in both classes. The delay in the onset of 
class~3 impacts by 1.4 days with respect to class~2 which was 
already recognized in the impact rate is easily seen: class~2 starts 
on day 174.0 and class~3 on day 175.4 (except for three rare 
earlier impacts which were detected during a spacecraft turn). 
When Galileo was approaching the inner 
Jovian system (days 174 to 177.5, cf. Fig.~\ref{c9orbit}), the 
particles were detected with $ \rm ROT = 270^{\circ} \pm 
70^{\circ}$. The distribution of class~2 particles over this range 
of 140$^{\circ}$ differs from that of class~3: between day 174.0 and 
day 175.4 most of the class~2 impacts are seen at rotation angles larger 
than 270$^{\circ}$. Furthermore, the region
$\rm ROT = 270^{\circ} \pm 10^{\circ}$ has 
very few class~2 impacts in this time span. Such a behaviour is not seen in 
the class~3 impacts after day 175.4: they do not show such a gap at 
$\rm ROT = 270^{\circ} \pm 10^{\circ}$.

In the inner Jovian system (day 178.0) the rotation angle of particles in 
both classes shifted by 180$^{\circ}$. This shift is coincident with 
the drop in the class~3 impact rate (see the top panel of Fig.~\ref{data}).
At this time the rotation angle for class~2 widened to nearly 
360$^{\circ}$ and the class~2 impact rate did not show a significant 
drop.

A closer look at the impacts with rotation angles of 
270$^{\circ} \pm 70^{\circ}$ reveals that the range of the 
rotation angles for class~2 is wider than that for class~3. This is illustrated 
in Fig.~\ref{c9rot}. The first panel shows the distribution of the 
rotation angles for all class~2 impacts detected between day 173.5 and 
day 177.5. 80\% of all particles are contained between 
the two dashed lines and the dash-dotted line shows the mean value.
The second panel shows the same histogram for class~3. Two things 
are obvious in the diagrams: a) the distribution of class~2 impacts 
is broader than that of class~3: 79$^{\circ}$ for class~2 vs. 
59$^{\circ}$ for class~3, and b) 
the distribution of class~2 shows a lack of particles with rotation 
angles smaller than 270$^{\circ}$.

To summarise, we have found four differences in the behaviour of the 
rotation angle between class~2 and class~3 impacts:

\begin{itemize}
\item[{\bf 1)}] The onset of class~3 impacts is delayed with respect to class~2,
\item[{\bf 2)}] The distribution of class~2 rotation angles is asymmetric,
\item[{\bf 3)}] The distribution of class~2 rotation angles is broader 
than that of class~3,
\item[{\bf 4)}] The gap in class~3 impacts during the shift by 
180$^{\circ}$ is `filled' with class~2.
\end{itemize}
 
\subsection{Analysis} \label{analysis}

We will now discuss one by one 
the four differences between class~2 and class~3 impacts listed above.

1) The delay of class~3 w.r.t class~2: the Jovian dust streams are  
collimated streams of nanometer-sized particles moving on curved trajectories 
outward from the inner Jovian system with velocities in excess of $\rm 300\,km\,s^{-1}$
(Zook et al. 1996, Gr\"un et al. 1998). Analysis of the angular momentum 
shows that particles on such trajectories can reproduce the onset of impacts, 
the $\rm 180^{\circ}$ shift in rotation angle   
when Galileo approaches Jupiter and the disappearance of impacts right 
after perijove passage (Gr\"un et al. in prep.). The time difference in the 
onset of class~2 and class~3 impacts can be attributed to different approach 
directions of the particles:

When Galileo is on the outer part of its orbit far away from Jupiter but 
moving towards the planet, the particles approach Galileo 'head-on' from 
the antenna direction, i.e. more or less parallel to the spacecraft spin 
axis (see also Gr\"un et al. 1998). Particles approaching from this 
direction cannot be detected by DDS (cf. Fig.~\ref{galileo}). When Galileo 
moves closer to Jupiter, the angle between the particles' approach direction and 
the spacecraft spin axis increases. At a certain position of Galileo, the 
particles begin to enter the FOV of the sensor (day 174.0 in the case of C9) 
and the first class~2 impacts occur. Later when the stream has moved further 
into the FOV the first class~3 impacts are detected (day 175.4). Still later when 
the particles approach from the anti-Earth direction the  
rotation angle   shifts by $\rm 180^{\circ}$ (day 178.0).

There is no generic difference between dust particles which can trigger 
class~2 or class~3 events. Apart from other charge rise time and coincidence 
criteria, three charge signals -- electron, ion and channeltron signal -- 
are required for a class~3 event, whereas only two are necessary for class~2
(Paper~IV). The delay in the onset of the class~3 impacts indicates that dust 
particles which approach the detector under large impact angles can trigger only 
class~2 events (cf. Fig.~\ref{galileo}). They do not create a channeltron 
signal and hence cannot create a class~3 event. Only 
particles approaching the detector under smaller impact angles can trigger 
channeltron signals. In other words, the sensitive area for class~3 
must be smaller than that for class~2. We will analyse the sensitive area and
the FOV for class~3 impacts in Sect.~\ref{DDSarea}

2) The asymmetry in the rotation angle distribution of class~2: this can be
understood as being due to shadowing caused by the magnetometer boom and 
the PLS and EPD instruments: particles which approach the sensor at rotation 
angles of $\rm 270^{\circ} \pm 10^{\circ}$ from the -Z direction and impact angles
between $\rm 40^{\circ}$ to $\rm 70^{\circ}$ hit the 
boom, EPD or PLS instead of the dust sensor. Hence, the boom 
is responsible for the lack of class~2 events at rotation angles of 
$\rm 270^{\circ}\pm 10^{\circ}$ in the lowest panel of Fig.~\ref{data}. On the other hand, 
particles which enter the sensor under such large impact angles
can only trigger class~2 events. Hence, the asymmetry in 
the rotation angle distribution is only seen in class~2. Some time 
later, when the stream has moved closer to the sensor axis, the dust 
particles can also trigger class~3 events:  only a small gap 
is seen in the rotation angle distribution of class~3 at about 270$^{\circ}$
(second panel of Fig.~\ref{c9rot}). 

3) The widths of the rotation angle distributions for both classes: 
the smaller width of the rotation angle distribution of class~3 compared 
with class~2 is consistent with a reduced sensitive area as discussed above:
during one spin revolution the stream sweeps across 
the FOV of DDS. Particles which enter the detector from 
close to the edge of the FOV
can trigger only class~2 events, whereas for class~3 the particles must 
be closer to the sensor axis. Hence, the rotation angle for a class~3 
detection must be closer to 270$^{\circ}$ than for a class~2 impact. 

4) The 'filling` of the gap in class~3 by class~2 particles: 
finally, to understand the class~2 impacts during the gap in class~3 
one has to consider the viewing geometry of DDS during one 
full spin revolution of Galileo. Since DDS is mounted at 60$^{\circ}$ 
with respect to the positive spin axis of the spacecraft and its FOV
is 70$^{\circ}$ half cone, particles approaching from within 10$^{\circ}$ of 
the positive spin axis can be detected at all rotation angles (Fig.~\ref{galileo}). 
Because of the consideration above, however, this is true only 
for class~2 impacts which were indeed detected with all rotation angles 
on day 178.0. At this time the dust stream approached Galileo from the 
anti-Earth direction. Due to the reduced sensitive area for 
particles which can trigger class~3 events and the corresponding reduced 
sensitive cone angle, class~3 events are missing at this time.

\section{The sensitive area for small class~3 dust impacts} \label{DDSarea}

The previous discussion has shown that the FOV of DDS for
class~3 impacts is significantly smaller than the usually adopted value 
of $70^{\circ}$ (Gr\"un et al. 1992a). The analysis, however, was only 
qualitative so far. The collimated Jovian dust streams (Gr\"un et al. 
1997, 1998) allow for an exact determination of the sensitive area 
for dust particles which can trigger class~3 events. 

We consider a subset of particles which approach the sensor with 
$0^{\circ}$ impact angle when the sensor points to a rotation angle of 
$\rm 270^{\circ}$. The ratio between the number of such
class~3 impacts, $\rm N_{3, \parallel}$, and the total number of 
such impacts (class~2 and class~3 together), $\rm N_{tot, \parallel}$, 
is proportional to the ratio between the sensitive area for 
class~3 impacts, $\rm A_{3}$, and the total sensitive area of the 
sensor, $\rm A_{tot}$: 
\begin{equation}\rm  \nonumber \frac{A_{3}}{A_{tot}} = 
                     \frac{N_{3, \parallel}}{N_{tot, \parallel}}
      \label{eqnarea}
\end{equation}
Because Galileo moves about Jupiter, the dust stream particles 
approach the sensor at rotation angles of
$\rm 270^{\circ}$ within several degrees along the sensor axis 
only at specific times. The times $\rm T_{||}$ when this 
occurs have been determined from a study of the angular momentum of the 
stream particles and their trajectories (Gr\"un et al. in prep). 
These times are given in Tab.~\ref{parsens}.
Due to the spacecraft spin (about three revolutions per minute), 
particles approach the sensor parallel to the sensor axis only during a
very short time interval, namely when the sensor points exactly
towards a rotation angle of $\rm 270^{\circ}$. Thus, the number of 
particles with $\rm ROT = 270^{\circ}$ is very small. In order to have 
sufficiently large numbers of particles, we take all dust impacts from 
a one day interval ($\pm 0.5$ days) around the times $\rm T_{||}$
given in Tab.~\ref{parsens} and a rotation angle interval of 
$\rm 20^{\circ}$ ($\rm 260^{\circ} \leq ROT \leq 280^{\circ}$).
We assume that the particles' approach direction is still close 
to the sensor axis in these time and rotation angle intervals.

Due to the low data transmission capability of Galileo, not all 
detected particles could be transmitted to Earth with their full 
information (cf. Sect.~\ref{denoising}). All particles, however, have been counted with 
one of 24 accumulators (Paper~I). For the particles which were only 
counted the rotation angle is unknown. The fraction of counted particles 
to that for which the full information was received on Earth depends on the 
class: class~3 is transmitted to Earth with a higher priority than class~2. To correct 
for this incomplete data transmission, a correction factor can be derived from the 
numbers of events counted with the accumulators of both classes for the 
lowest amplitude range (AR1), 
$\rm N_{2,acc}$ and $\rm N_{3,acc}$. The accumulators give the numbers 
of events independent of the rotation angle, and the class~2 accumulator
also contains noise events. Assuming that the full information 
of an event is transmitted to Earth with the same priority 
independent of the rotation angle, the number derived from the 
accumulator can simply be scaled with the ratio between the number 
of particles detected with $\rm 260^{\circ} \leq ROT \leq 280^{\circ}$, 
$\rm N_{x,260-280}$ ($\rm x = 2,3$), and the number of 
particles detected over the full rotation angle range 
($\rm 0^{\circ} \leq ROT \leq 360^{\circ}$),
$\rm N_{x,0-360}$:
\begin{eqnarray*}  \rm  
    N_{3,\parallel}&\rm =& \rm N_{3,acc}\cdot\frac{N_{3,260-280}}{N_{3,0-360}} \\
\rm   N_{tot,\parallel}&\rm 
           = & \rm (N_{2,acc}\cdot (1-f_{noi}) + N_{3,acc}) \cdot 
       \frac{N_{2,260-280} + N_{3,260-280}}{N_{2,0-360} + N_{3,0-360}} \\
\end{eqnarray*}
For the number of events derived from the class~2 accumulator, $\rm N_{2,acc}$,
only the fraction $\rm 1 - f_{noi}$ has been taken into account 
(cf. Sect.~\ref{denoising}). Table~\ref{parsens} lists the numbers required to 
calculate the sensitive area of the dust sensor for small class~3 impacts. Only 
data from 1997 have been used. In earlier data sets unknown numbers of accumulator
overflows occurred, especially in class~2, which makes the determination of 
$\rm N_{2,acc}$ unreliable for those orbits. The mean ratio of the sensitive areas 
derived from the seven 1997 orbits is $\rm A_3/A_{tot} = 0.11\,\pm\,0.04$. Given the 
total sensitive area of $\rm A_{tot} = 1,000\,cm^2$ (Gr\"un et al. 1992a), the 
sensitive area for class~3 impacts becomes $\rm A_3=110\,\pm\, 37\,cm^2$.

The sensitive area of the Galileo dust sensor as a function of impact angle 
has been published by Gr\"un et al. (1992a) for a total sensitive 
area of $\rm A_{tot} = 1,000\, cm^2$. The corresponding sensor opening angle is 
$\rm 70^{\circ}$. The above newly determined sensitive area for small class~3 
impacts allows for a new calculation of this function (Fig.~\ref{sensareadds}). 
A sensitive area of $\rm 110\,cm^2$ for particles approaching the sensor 
parallel to the sensor axis has been adopted. Due to shielding by the 
channeltron housing the sensitive area is a ring with the channeltron in its 
center (cf. Fig.~\ref{ddsconf}). Starting from an initial value of 
$\rm 110\,cm^2$ the sensitive area rises with increasing impact 
angle and reaches a maximum of $\rm 160\,cm^2$ for impact angles close to  
$\rm 27^{\circ}$. The sensitive area for impact angles of $\rm 0^{\circ}$ 
is smaller than that for larger angles because the channeltron housing 
shields a larger area for smaller angles.
(Fig.~\ref{ddsconf}). For particles approaching 
at impact angles significantly larger than $\rm 27^{\circ}$ the sensitive area 
drops because shielding by the sensor's side walls becomes important. At an impact 
angle of $\rm 48^{\circ}$ the sensitivity drops to zero. A special case 
are particles which approach the sensor from within about $\pm 10^{\circ}$
of the -Z direction where shielding of the boom becomes important 
(Fig.~\ref{ddsfov}): for impact angles larger than $\rm 33^{\circ}$ the 
sensitivity drops to zero because particles hit the boom instead of the 
sensor target.  The width of the region where no class~2 particles 
have been detected with rotation angles $\rm 270^{\circ} \pm 10^{\circ}$ 
from day 174.0 to day 175.4 (Fig.~\ref{data}, Sect.~\ref{analysis} point 2) 
reflects the width of the boom (Fig.~\ref{ddsfov}).

The angular sensitivity averaged over one spin revolution of Galileo can be 
found in Gr\"un et al. (1992b, note that a mounting angle of 55$^{\circ}$ 
has been used in that analysis). This calculation is repeated in 
Fig.~\ref{sensarea}, for a mounting angle of 60$^{\circ}$. The upper curve has 
been calculated with a sensitive area of the dust sensor of $\rm 1,000\, cm^2$ which 
corresponds to a FOV of $\rm 140^{\circ}$. The lower curve shows 
the spin-averaged sensitivity for a  sensitive area of $\rm 110\, cm^2$. The 
spin-averaged FOV is now reduced to $\rm 105^{\circ}$ and has a gap 
between $\rm 0^{\circ}$ and $\rm 12^{\circ}$. The upper curve in Fig.~\ref{sensarea} 
explains the widening of the rotation angle of the class~2 impacts on day 178.0 
and the bottom panel explains why the class~3 impacts show a 'gap' at the same 
time: with the reduced sensitive area for small class~3 impacts no particles 
approaching from 
the direction parallel to the positive spin axis (anti-Earth direction) can be 
detected anymore. The spin averaged sensitive area for small class~3 impacts has 
a maximum of $\rm 35\, cm^2$ which is a factor of 7 smaller than the maximum of the 
total sensitive area ($\rm 235\, cm^2$, compare the lower and the upper curve in 
Fig.~\ref{sensarea}). 
This is in good 
agreement with the ratio between the class~3 and the total impact rates of stream
particles detected during the orbits E6 to E12 (not only the short time intervals 
considered in Tab.~\ref{parsens}).

\section{Summary and Conclusions}

Galileo dust data obtained within about $\rm 20\,R_J$ from Jupiter show clear 
indications of noise contamination in their second highest event class (class~2). 
The highest event class (class~3) is almost totally noise-free. We have derived 
criteria 
to reject the noise events in class~2 and obtained a denoised set of dust data.
This set of class~2 and class~3 dust impacts is now considered as the 
complete set of dust data from Galileo's prime Jupiter mission. The present work 
extends the number of dust particles identified in the Jovian system by almost a 
factor of 2 with respect to earlier works which could use only class~3 
(e.\,g. Gr\"un et al. 1998). Since, on average, class~2 impacts of stream particles 
have been detected further away from Jupiter than class~3 impacts, the space sampled by 
the Galileo dust measurements in the Jovian system is extended with
the new data set. 

Investigations of dust in the Jupiter system will benefit from the new extended 
data set in various ways.
Time series analysis relies on data sets
for long time periods. Presently ongoing work which aims at the determination
of the ultimate source of the Jupiter dust streams (A. Graps et al. in prep.) 
significantly benefits from the extended data set. 
Impact rates of 
stream particles are now about a factor of 9 higher than reported earlier 
(Gr\"un et al. 1998, which used only class~3 particles). A few
impacts have been detected in the vicinity of the Galilean satellites during
close flybys at these moons which have been created by impacts of other dust 
particles onto the surfaces of the moons. Only with the extended noise-free data 
set a detailed investigation of the spatial distribution of these dust grains
has become possible (Kr\"uger et al. 1999). 

The nanometer-sized Jupiter streams
have been used to analyse the sensitive area and the FOV  of 
the detector itself.
Small dust particles which approach the sensor under large impact angles cannot 
create an impact plasma cloud which is large enough to trigger a channeltron 
signal. The sensitive area for small class~3 dust impacts is 
$\rm 110\,\pm 37\, cm^2$ which is almost a factor of 10 smaller than
the total sensitive area of the detector which is valid for class~2 ($\rm 1,000\,cm^2$). 
The corresponding 
FOV is $\rm 96^{\circ}$ instead of $\rm 140^{\circ}$.
Shadowing of the sensor FOV by the magnetometer boom, the  Energetic 
Particles Detector and the Plasma Subsystem is recognized  in the data.

The sensitive area of the dust sensor averaged over one spacecraft spin 
revolution has been calculated for the reduced FOV.
Because the dust sensor is mounted with an angle of 
$\rm 60^{\circ}$ to the spacecraft spin axis and the sensor FOV
is reduced to only $\rm 48^{\circ}$ half cone angle, small particles 
approaching the sensor to within $\rm 12^{\circ}$ parallel to the spacecraft 
spin axis cannot trigger class~3 events.

Only the very fast ($\rm v > 300 \, km\, s^{-1}$)
nanometer-sized stream particles have been used for our analysis of the FOV because 
they approached the sensor as 
collimated streams. With particles approaching the sensor from a wide range 
of directions the reduced FOV would have been hardly recognizable. 
According to our present understanding, the reduction of the sensitive area for 
class~3 applies only to these very fast and tiny 
particles which originate from the Jupiter system. They create impact charge 
signals in amplitude range AR1 only.
We do not expect a significant reduction of the sensitive area for
bigger particles which create charge signals in higher amplitude                            
ranges (AR2 to AR6). 
Even particles which are bigger and slower than the stream particles
($\rm v < 70 \, km\, s^{-1}$) and which are recognized as 
class~3 impacts should ''see`` the total unreduced sensitive area.

Thus, conclusions obtained earlier about interplanetary 
dust (Gr\"un et al. 1997) and interstellar dust (Landgraf 1998) 
are not affected by the reduced FOV for class~3. Also results about
the dust populations in the Jupiter system other than the stream
particles (Gr\"un el al. 1998) remain 
unchanged.
Furthermore, results about the Jupiter dust streams obtained  from 
measurements in interplanetary space (Gr\"un et al., 1993, 1996) 
remain unaffected because they used (among other parameters) the
average impact direction rather than the widths of the streams.

The reduced FOV for class~3 can be used for an estimate of the 
dispersion of the dust streams which approach the detector from a 
source close to Jupiter. 
From the time difference between the onset of class~2 and class~3 
impacts and the shift of the
impact angle during this time interval we can estimate how strongly 
collimated the streams are. In the case of C9 the time difference is
1.4 days and the shift in impact angle in this time interval is $\rm 22^{\circ}$
(which is the difference between the FOV for class~2 and class~3).
From Fig.~\ref{data} it is evident that within a few hours the 
impact rate increased from a very low background level to about one 
impact within 10 minutes. This implies that the width of the dust 
streams is only a few degrees and it is consistent with 
theoretical considerations (Horanyi et al. 1997, Gr\"un et al. 1998).

A drop in the class~3 impact rate has been repeatedly observed 
when Galileo was approaching Jupiter during 
its orbital tour about the planet (Gr\"un el al. 1998). This drop 
occurred 
around the times when the particles seemed to approach from all rotation angles.
Formerly, it could be only partially explained by a dawn-dusk asymmetry in the 
release of dust particles from the Io torus (Horanyi et al. 1997; Heck 1998).
The reduction in the sensitive area of the dust instrument found in 
the present analysis nicely explains the drop in the class~3 impact rate
which is not seen in class~2.

In an analysis of the angular momentum of the stream particles
Gr\"un et al. (in prep.) constrain the possible distance range 
of the source from Jupiter. This 
analysis relies on the impact direction of the particles 
onto the detector (impact angle). Agreement 
with the onset of the class~3 impacts could only be achieved with
the reduced sensitive area. 

It should be noted that our analysis of the sensor FOV could not 
have been done during 
ground calibration in the laboratory. The stream particles are much 
faster and smaller than 
particle speeds and masses achieveable with the dust accelerator 
(Gr\"un et al. 1992a).

\hspace{1cm}

{\bf Acknowledgments.}
The authors appreciate valuable discussions with Kai-Uwe Thiessenhusen 
during the preparation of the manuscript. We thank the referees, 
D. P. Hamilton and J. A. M. McDonnell, for valuable suggestions which 
improved the presentation of our results. 
We also thank the Galileo project at JPL for effective and successful 
mission operations. This work has been supported by Deutsches
Zentrum f\"ur Luft- und Raumfahrt e.V. (DLR).

\section*{References}

{\small

{\bf Baguhl, M., Gr\"un, E., Linkert, D., Linkert, G.\ and Siddique, N.,}
Identification of 'small' dust impacts in the Ulysses dust
detector data. {\em  Planet.\ Space Sci. }{\bf 41}, 1085-1098, 1993

{\bf Frank, L. A., Ackerson, K. L., Lee, J. A., English, M. R. and Pickett, G. L.}
The Plasma Instrumentation for the Galileo mission. 
{\em Space Sci.\ Rev. }{\bf 60}, 283-307, 1992


{\bf Gr\"un, E., Fechtig, H., Hanner, M.S., Kissel, J., Lindblad, B-A.,
Linkert, D., Linkert, G., Morfill, G.E.\ and Zook, H.A.,}
The Galileo Dust Detector. {\em Space Sci.\ Rev. }{\bf 60}, 317-340, 1992a

{\bf Gr\"un, E., Fechtig, H., Giese, R.H., Kissel, J., Linkert, D.,
Maas, D., McDonnell, J.A.M., Morfill, G.E., Schwehm, G.\ and
Zook, H.A.,} The Ulysses dust experiment.
{\em Astron.\ Astrophys.\ Suppl.\ Ser. }{\bf 92}, 411-423, 1992b

{\bf Gr\"un, E., Zook, H.A., Baguhl, M., Balogh, A., Bame, S.J.,
Fechtig, H., Forsyth, R., Hanner, M.S., Horanyi, M., Kissel, J.,
Lindblad, B.-A., Linkert, D., Linkert, G., Mann, I., McDonnell, J.A.M.,
Morfill, G.E., Phillips, J.L., Polanskey, C., Schwehm, G., Siddique, N.,
Staubach, P., Svestka, J. and Taylor, A., } Discovery of jovian
dust streams and interstellar grains by the Ulysses spacecraft. {\em
Nature }{\bf 362}, 428-430, 1993

{\bf Gr\"un, E., Baguhl, M., Fechtig, H., Hamilton, D.P., Kissel, J.,
Linkert, D., Linkert, G.\ and Riemann, R.,}
Reduction of Galileo and Ulysses dust data.
{\em Planet. Space Sci.} {\bf 43}, 941-951, 1995a (Paper I)

{\bf Gr\"un, E., Baguhl, M., Divine, N., Fechtig, H., Hamilton, D. P.,
Hanner, M. S., Kissel, J., Lindblad, B.-A., Linkert, D., Linkert, G., 
Mann, I., McDonnell, J. A. M., Morfill, G. E., Polanskey, C., Riemann, R.,
Schwehm, G., Siddique, N., Staubach P. and Zook, H. A.,}
Three years of Galileo dust data. {\em Planet. Space Sci.}, {\bf 43}, 
953-969, 1995b (Paper II)

{\bf Gr\"un, E., Baguhl, M., Divine, N., Fechtig, H., Hamilton, D.P.,
Hanner, M.S., Kissel, J., Lindblad, B.-A., Linkert, D., Linkert, G.,
Mann, I., McDonnell, J.A.M., Morfill, G.E., Polanskey, C.,
Riemann, R., Schwehm, G., Siddique, N., Staubach, P.\ and Zook,
H.A., } Two years of Ulysses dust data, {\em Planet. Space Sci.} 
{\bf 43}, 971-999, 1995c (Paper III)

{\bf Gr\"un, E.,
Baguhl, M., Hamilton, D. P., Riemann, R., Zook, H. A.,
Dermott, S., Fechtig, H., Gustafson, B. A., Hanner, M. S., Hor\'anyi, M.,
Khurana, K. K., Kissel, J., Kivelson, M., Lindblad, B.-A., Linkert, D.,
Linkert, G., Mann, I., McDonnell, J. A. M., Morfill, G. E., Polanskey, C.,
Schwehm, G., and Srama, R.,}
Constraints from Galileo observations on the origin of jovian dust
streams,
{\em Nature}, 381, 395-398, 1996.

{\bf Gr\"un, E., Kr\"uger, H., Dermott, S., Fechtig, H., Graps, A., 
Gustafson, B. A., Hamilton, D. P., Hanner, M. S., Heck, A., 
Hor\'anyi, M., Kissel, J., Lindblad, B.-A., Linkert, D., Linkert, 
G., Mann, I., McDonnell, J. A. M., Morfill, G. E., Polanskey, C., 
Schwehm, G., Srama, R. and Zook, H. A.}
Dust measurements in the jovian magnetosphere. {\em Geophys. Res. Lett.}
{\bf 24}, 2171-2174, 1997


{\bf Gr\"un, E., Kr\"uger, H., Graps, A., Hamilton, D. P., Heck, A.,
Linkert, G., Zook, H. A., Dermott, S., Fechtig, H., Gustafson, B. A.,
Hanner, M. S., Hor\'anyi, M., Kissel, J., Lindblad, B.-A., Linkert, D.,
Mann, I., McDonnell, J. A. M., Morfill, G. E., Polanskey, C., 
Schwehm, G., Srama, R.}  Galileo Observes Electromagnetically 
Coupled Dust in the Jovian Magnetosphere. {\em J. Geophys. Res.}, 
{\bf 103}, 20011-20022, 1998

{\bf Heck, A.,} Modellierung und Analyse der von der Raumsonde 
Galileo im Jupitersystem vorgefundenen Mikrometeoroiden-Populationen,
PhD thesis, Heidelberg, 1998

{\bf Horanyi, M., Gr\"un, E., Heck, A.} Modelling the Galileo
dust measurements at Jupiter, {\em Geophys. Res. Lett.}, {\bf 24}, 
2175-2178, 1997

{\bf Kivelson, M. G., Khurana, K. K., Means, J. D., Russell, C. T. and
Snare, R. C.} The Galileo magnetic field investigation 
{\em Space Sci.\ Rev. }{\bf 60}, 357-383, 1992

{\bf Kr\"uger, H.,  Gr\"un, E., Hamilton, D. P., Baguhl, M., Dermott, 
S., Fechtig, H., Gustafson, B. A., Hanner, M. S., Heck, A.,
Hor\'anyi, M., Kissel, J., Lindblad, B.-A., Linkert, D., Linkert, G.,
Mann, I., McDonnell, J. A. M., Morfill, G. E., Polanskey, C., 
Riemann, R., Schwehm, G., Srama, R. and Zook, H. A., }
Three years of Galileo dust data: II. 1993 to 1995. 
{\em Planet. Space Sci.}, 1998a, in press (Paper~IV) 

{\bf Kr\"uger, H.,  Gr\"un, E., Graps, A. and Lammers, S.} Observations
of electromagnetically coupled dust in the Jovian magnetosphere,
{\em Astrophys. and Space Sci.}, in press, 1998b

{\bf Kr\"uger, H., Krivov, A. V., Hamilton, D. P., Gr\"un, E.,} 
Discovery of a dust cloud around Ganymede, {\em Nature}, 1999, submitted

{\bf Landgraf, M.} Modellierung der Dynamik und Interpretation der 
In-Situ-Messung interstellaren Staubs in der lokalen Umgebung des
Sonnensystems, PhD. thesis, Heidelberg, 1998

{\bf Williams, D. J., McEntire, R. W., Jaskulek, S and Wilken, B.}
The Galileo Energetic Particles Detector. 
{\em Space Sci.\ Rev. }{\bf 60}, 385-412, 1992

{\bf Zook, H. A., Gr\"un, E., Baguhl, M., Hamilton, D. P., 
Linkert, G., Liou, J.-C., Forsyth, R., Phillips, J. L. }
Solar wind magnetic field bending of jovian dust trajectories. 
{\em Science} {\bf 274}, 1501-1503, 1996

\clearpage


\begin{table}[hb]
\caption{\label{criteria}
Criteria for separation of class~2 noise events 
from dust impacts in the inner Jovian system. Noise 
events in the lowest amplitude range (AR1) fulfill at least one 
of the criteria listed below 
(see Paper~I for a definition of the parameters), whereas 
noise events in the higher amplitude ranges fulfil all criteria 
listed for AR2 to AR6.
}
{\small
  \begin{tabular*}{15cm}{lcc}
   \hline
   \hline \\[-2.5ex]
Charge Parameter  &  AR1                         & AR2 to AR6   \\
\hline  
Entrance grid amplitude                  &  PA $\geq$ 9                    &    ---       \\
Channeltron amplitude                    &  ---                            & CA $\leq$ 2  \\
Target amplitude minus iongrid amplitude &  (EA$-$IA) $\leq$ 1 or          & (EA$-$IA) $\leq$ 1 or \\ 
                                         &  (EA$-$IA) $\geq$ 7             & (EA$-$IA) $\geq$ 7 \\
EA risetime                                      & ET $\leq$ 9 or ET $\geq$ 14     &    ---       \\
IA risetime                                      & IT $\leq$ 8                     &    ---       \\
Flighttime target - iongrid                      & EIT $\leq$ 3 or EIT $\geq$ 14   &    ---       \\
Flighttime entrance grid - target                & 1 $\leq$ PET $\leq$ 30          &    ---       \\
\hline
\hline
\end{tabular*}\\[1.5ex]
}
\end{table}

\begin{table}[hbt]
\caption{\label{eventnumber}
Total number of class~2 noise events and class~2 and class~3 
dust impacts determined from two years of Galileo dust data 
(1996 and 1997, G1 to E12 orbits)
for the six ion amplitude ranges AR. 
}
  \begin{tabular}{ccccc}
   \hline
   \hline \\[-2.5ex]
      AR  &   CL2       &  CL2      & CL3      & Sum     \\
          &       Noise &      Dust &     Dust &     Dust\\
\hline
       1  &    2,690    &   4,547   &  5,864   & 10,411\\
       2  &      128    &      85   &     51   &    136\\
       3  &        3    &      27   &     35   &     62\\
       4  &        1    &      16   &     17   &     33\\
       5  &        0    &       7   &      8   &     15\\
       6  &        0    &       3   &      0   &      3\\
\hline
 Sum      &    2,822    &   4,685   &  5,975   & 10,660\\
\hline
\hline
\end{tabular}\\[1.5ex]
\end{table}

\begin{table}[hb]
\caption{\label{parsens}
Data used for the determination of the sensitive area of DDS
for class~3 particles. Column (1) gives the orbit number and column (2) the 
times $\rm T_{||}$ when the dust streams approached the sensor closest to the 
sensor axis at rotation angles of $270^{\circ}$. Columns (3) to (6) give the 
numbers of class~2 and class~3 particles for rotation angle ranges $\rm 
260^{\circ}$ to $280^{\circ}$ and $\rm 0^{\circ}$ to $ 360^{\circ}$, respectively. 
Columns (7) and (8) give the numbers of events determined from the accumulators 
(i.e. dust impacts plus noise for class~2, dust impacts for class~3).
Column (9) lists the noise fraction in class~2 and, finally, column (10)
gives the ratio between the sensitive area for class~3 impacts and the total
sensitive area determined from the data of the specific encounter.
}
\footnotesize
  \begin{tabular}{ccccccccccc}
   \hline
   \hline \\[-2.5ex]
 Orb. & $\rm T_{||}$ & $\rm N_{2,260-280}$ & $\rm N_{3,260-280}$ & $\rm N_{2,0-360}$ & $\rm N_{3,0-360}$ & 
 $\rm N_{2,acc}$ & $\rm N_{3,acc}$ & $\rm f_{noi}$ & $\frac{\rm A_3}{\rm A_{tot}}$ \\[0.5ex]
       & [year-doy] &     &    &    &     &       &       &     & \\
  (1)  &   (2)      &        (3)      & (4)&(5) &(6) & \multicolumn{1}{c}{(7)} & 
\multicolumn{1}{c}{(8)}  &  \multicolumn{1}{c}{(9)} & (10) \\
\hline
   E6  &  97-049.69 &  31 & 10 & 84 & 35  & 302   &   40  & 0.10& 0.11\\
   G7  &  97-092.69 &  7  & 46 & 21 & 210 & 5,614 &  953  & 0.09& 0.15\\
   G8  &  97-127.03 &  7  & 50 & 17 & 246 & 12,199& 1,830 & 0.26& 0.16\\
   C9  &  97-177.10 &  25 & 27 & 96 & 111 &  3,290&   298 & 0.04& 0.08\\
   C10 &  97-260.33 &  21 & 28 & 109& 132 &  5,487&   524 & 0.22& 0.11\\
   E11 &  97-309.42 &  30 & 15 & 105&  75 &  2,525&   246 & 0.06& 0.08\\
   E12 &  97-348.77 &  8  & 10 & 37 &  34 &  1,827&    90 & 0.05& 0.06\\
\hline
\hline
\end{tabular}\\[1.5ex]
\end{table}

\clearpage


\begin{figure}[h]
\epsfxsize=8.2cm
\epsfbox{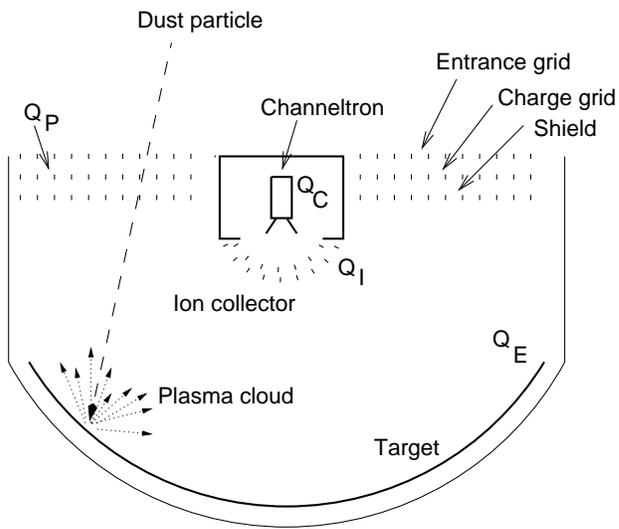}
        \caption{\label{ddsconf}
Schematic configuration of the Galileo dust detector (DDS).
Up to three charge signals ($\rm Q_I, Q_E, Q_C$) are used for dust impact 
identification.
}
\end{figure}

\begin{figure}
\epsfxsize=12.0cm
\epsfbox[-60 300 320 750]{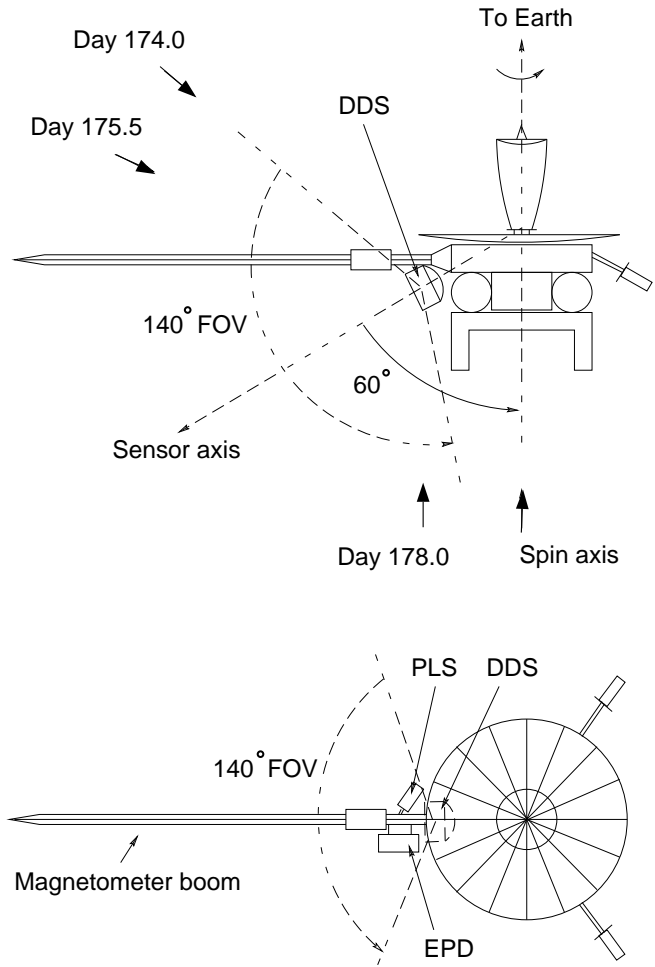}
        \caption{\label{galileo} Galileo spacecraft configuration (schematic); 
top: side view; bottom: top view. DDS is mounted underneath the magnetometer 
boom and the sensor field of view (FOV) is indicated. The approach 
directions of the dust stream particles for three times (days 174.0, 175.4 and
178.0) discussed in the text are marked by arrows.
}
\end{figure}

\begin{figure}
\epsfxsize=10.5cm
\epsfbox{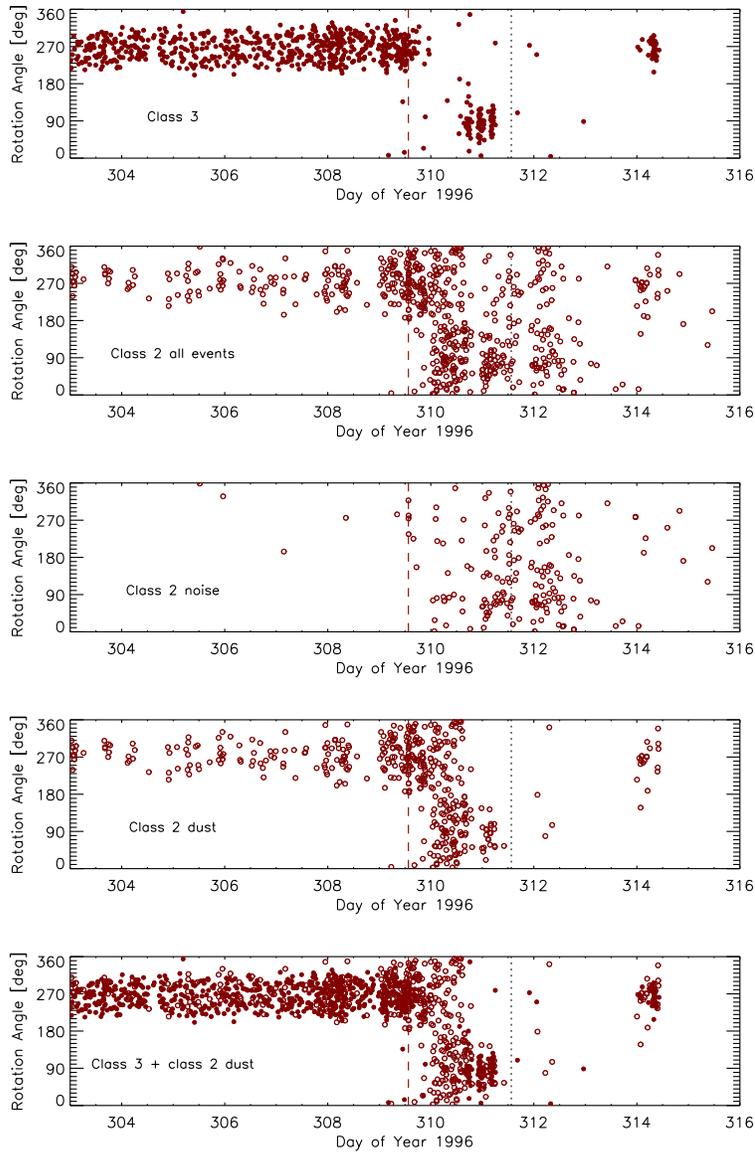}
        \caption{\label{noise1}
Rotation angle for class 3 and class 2 events in the lowest 
ion amplitude range at Galileo's 
C3 encounter. First panel: class 3 dust impacts; 
second panel: all class 2 events detected; third panel: 
noise events recognised; fourth panel: class 2 dust impacts
after denoising; fifth panel: class 2 and 
class 3 dust impacts. Class 3 is always shown as filled 
circles and class 2 as open circles. Callisto closest approach is 
indicated by a dashed line, perijove passage by a 
dotted line.
}
\end{figure}

\begin{figure}
\epsfxsize=11.5cm
\epsfbox{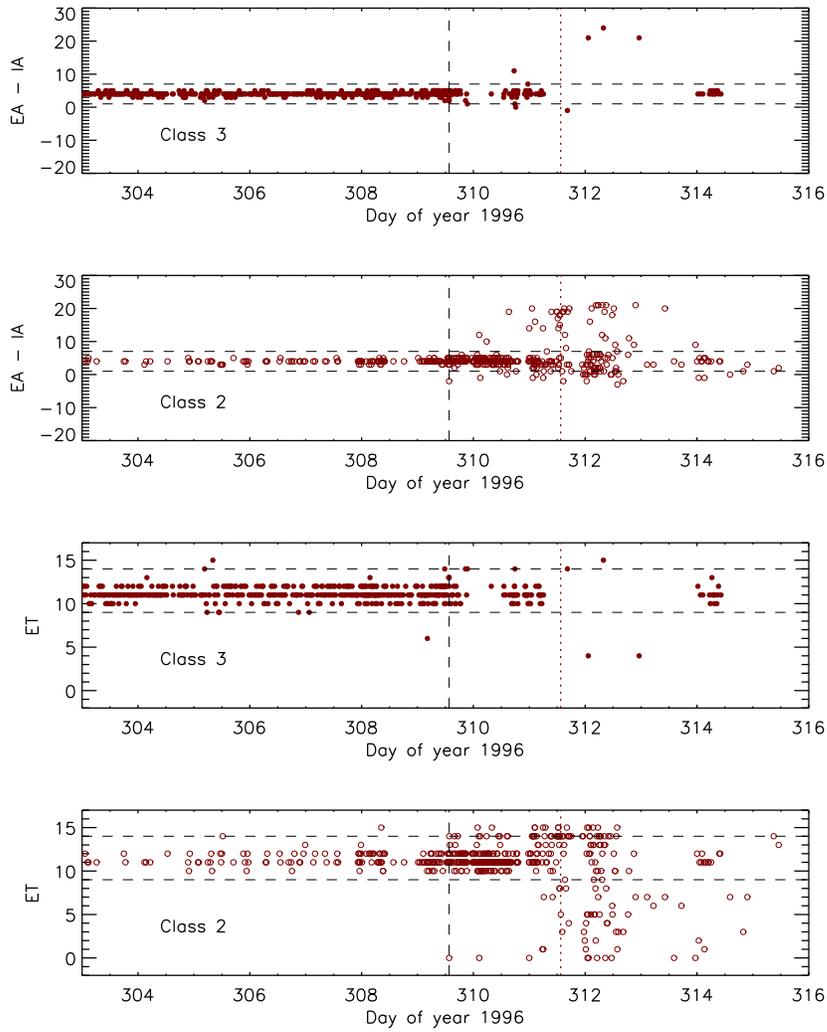}
        \caption{\label{noise2}
Impact parameters determined for the dust impacts shown in 
Fig.~\ref{noise1}: Upper two panels: digital value of 
charge measured on the target minus digital value of charge 
measured on the ion collector grid. Lower two panels: 
rise time of charge signal measured on the target (digital value). 
In both sets the upper panels show class 3 data 
and the lower ones class 2 data. Callisto closest approach is
indicated by a dashed line, perijove passage by a
dotted line. The horizontal lines indicate the range of the 
parameters used to separate noise events from dust impacts.
}
\end{figure}

\begin{figure}
\epsfxsize=14.5cm
\epsfbox{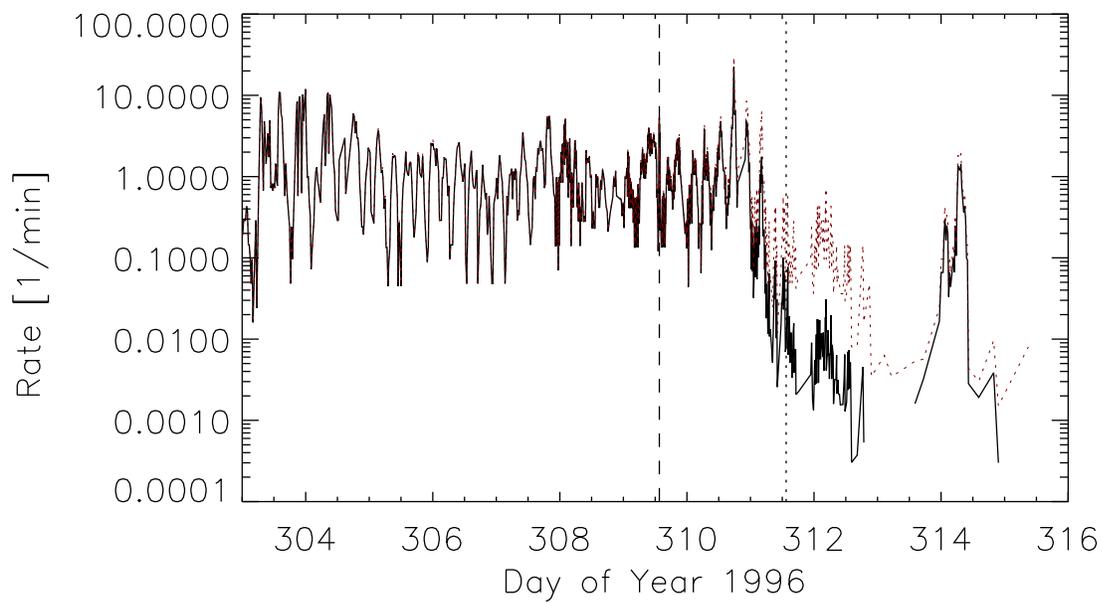}
        \caption{\label{noise3}
Class 2 rate before denoising (dotted line) and after denoising 
(solid line). Vertical lines indicate Callisto closest approach (dashes)
and perijove passage (dots).
}
\end{figure}

\begin{figure}
\epsfxsize=14.2cm
\epsfbox{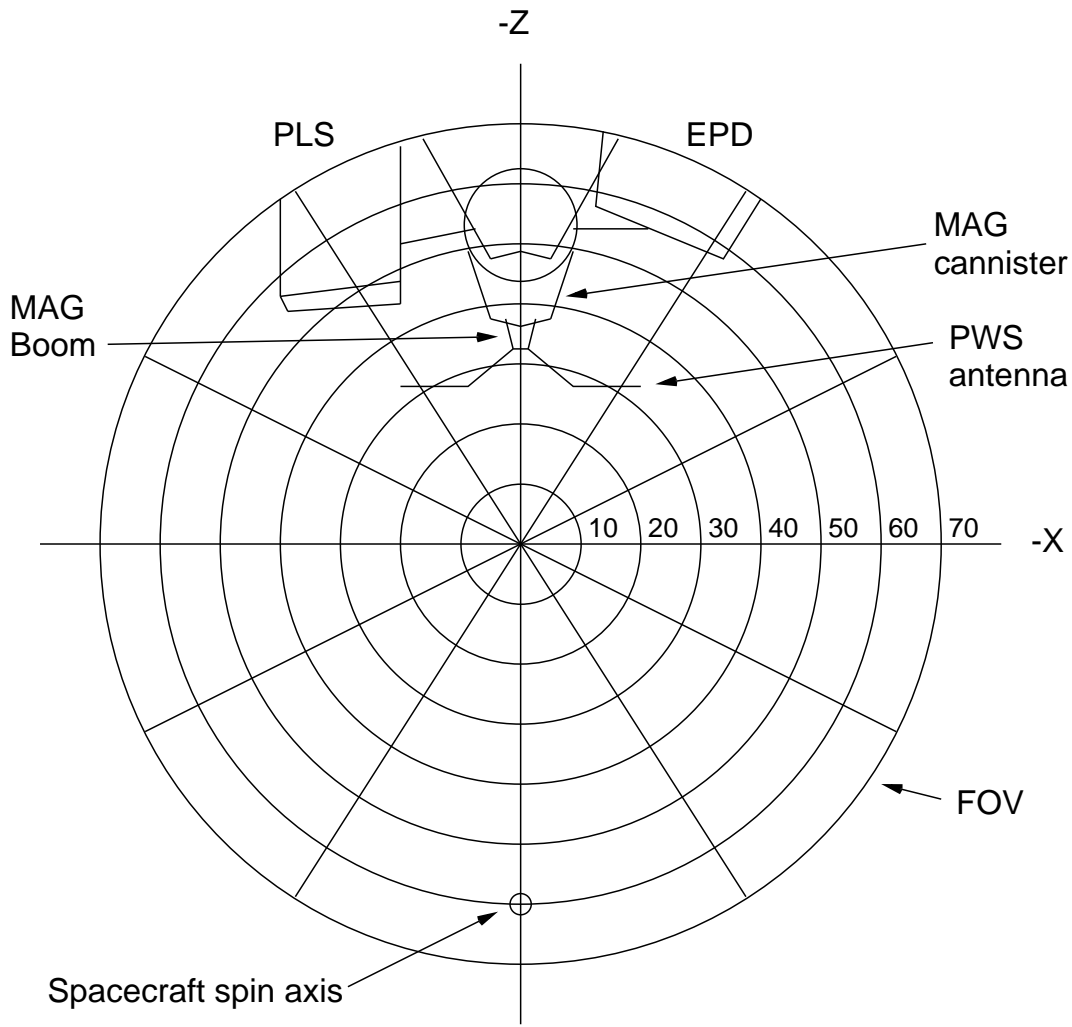}
        \caption{\label{ddsfov}
DDS field of view (FOV) and obscuration by the MAG, PLS 
and EPD instruments as seen from a point on the target on the sensor axis (from
JPL Galileo document GLL-3-180C). Numbers give the angle w.r.t. the 
sensor axis (impact angle). The spacecraft spin axis is towards 
$\rm 60^{\circ}$ in +Z direction. During one spin revolution of the 
spacecraft a stream of particles approaching under a given impact angle
moves through the FOV along an arc centered on the spin axis. 
}
\end{figure}

\begin{figure}
\epsfxsize=9.9cm
\epsfbox{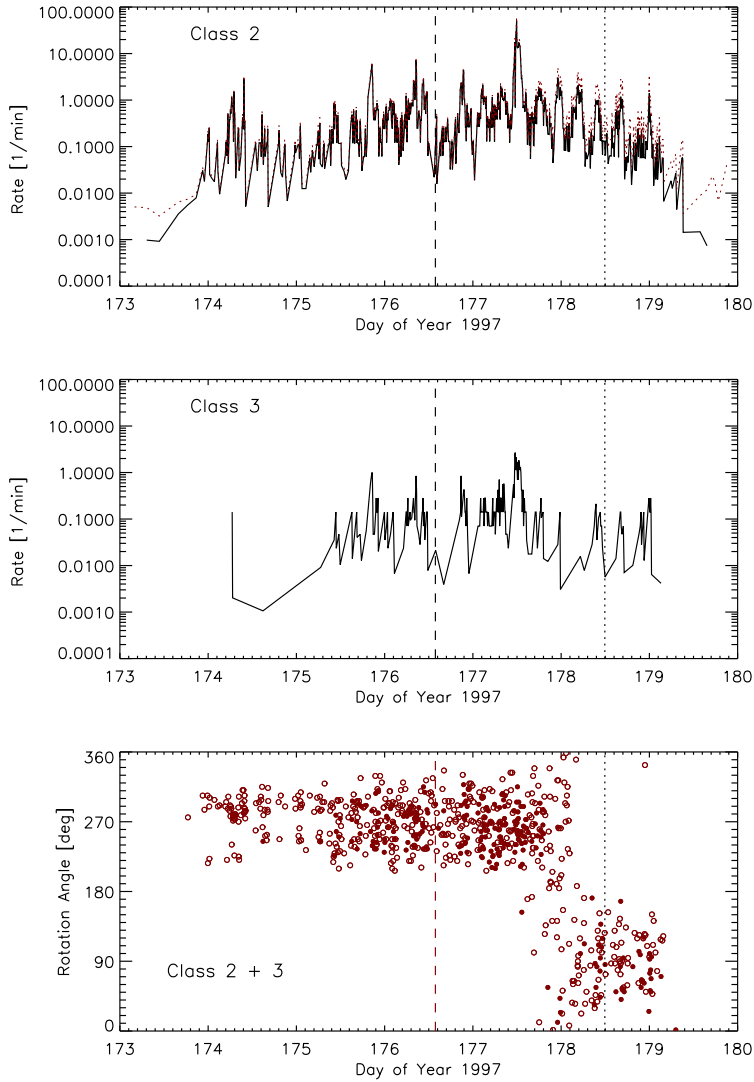}
        \caption{\label{data} Impact rate and rotation angle of 
dust particles in the lowest amplitude range (AR1) for classes 2 and 3 
as a function of time during Galileo's C9 orbit. The 
dashed vertical line indicates the C9 closest approach, and the dotted line 
indicates perijove passage. Top panel: 
class~2 impact rate corrected for noise in the inner Jovian system (solid curve) 
and total event rate without denoising (dashed curve, best visible after day 
178); middle panel: class~3 impact rate, the spike on day 174.2 is caused by 
a few impacts detected during a spacecraft turn;
bottom panel: rotation angle of class~2 (open circles, denoised) and 
class~3 (filled circles) dust impacts.  
}
\end{figure}

\begin{figure}
\epsfxsize=11.5cm
\epsfbox{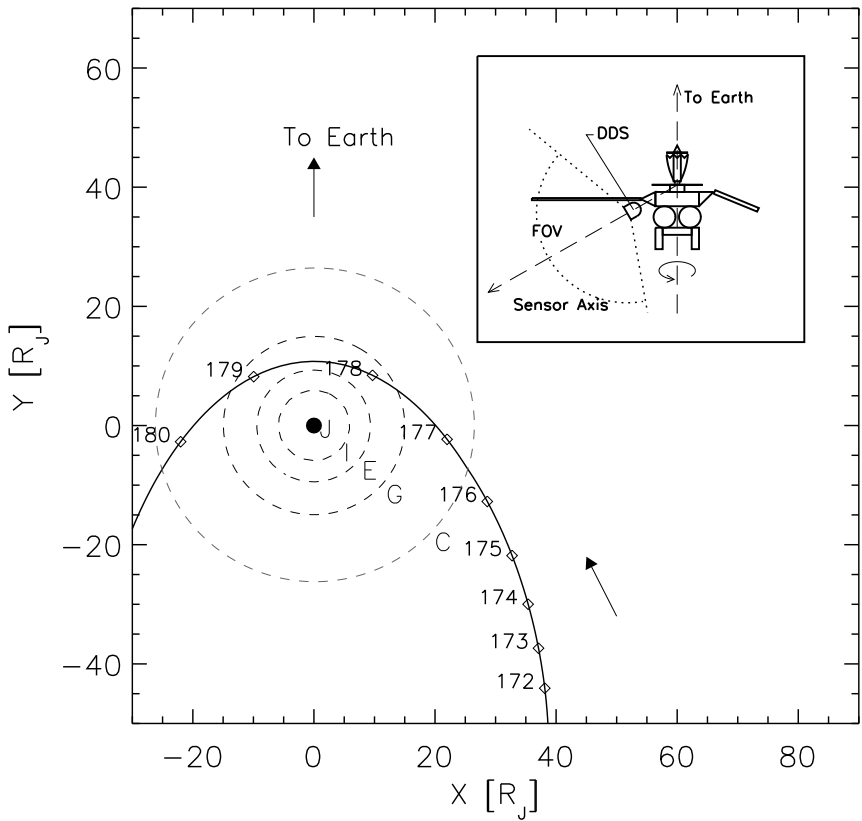}
\caption{\label{c9orbit} 
Geometry of Galileo's C9 orbit (solid line). The orbits of the Galilean
satellites are shown as dashed lines. Squares indicate Galileo's position at 
the beginning of each day of year 1997. Earth is to the top. The inset shows 
the orientation of DDS for a rotation angle of $270^{\circ}$.
}
\end{figure}

\begin{figure}
\epsfxsize=13.9cm
\epsfbox{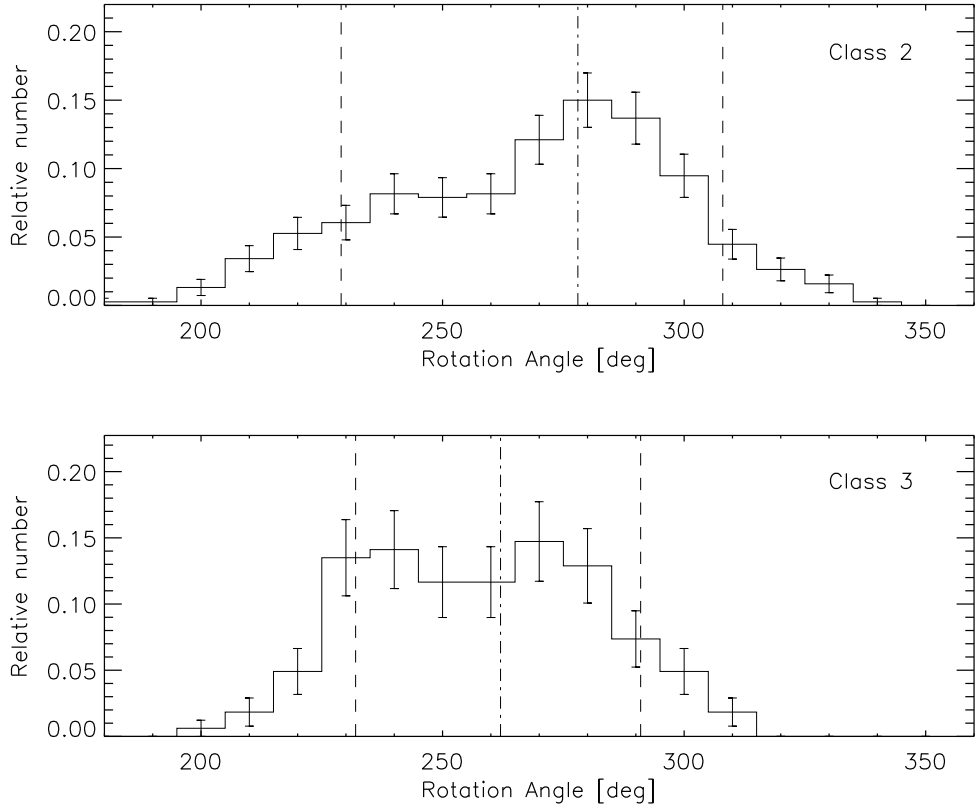}
        \caption{\label{c9rot}
Rotation angle distribution for class~2 (upper panel) and class~3 
(lower panel) dust impacts in the lowest amplitude range (AR1) detected 
between day 173.5 and day 177.5 (C9 orbit). For both classes the 
relative number of impacts per rotation angle bin is shown. 
The dash-dotted line indicates the mean rotation angle 
($\rm 278^{\circ}$ for class~2 and $\rm 262^{\circ}$ for 
class~3, respectively) and the 
dashed lines show the range which contains 80\% of all particles. 
The width of this 80\% range is 79$^{\circ}$ for class~2 
and 59$^{\circ}$ for class~3, respectively.
}
\end{figure}

\begin{figure}
\epsfxsize=11.5cm
\epsfbox{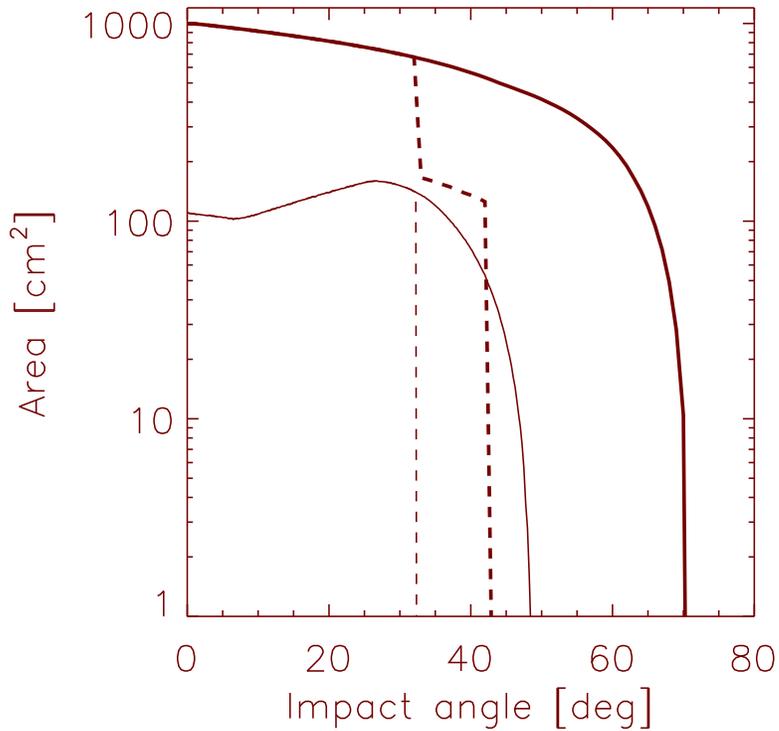}
        \caption{\label{sensareadds}
Sensitive area of DDS as a function of angle w.r.t. the 
sensor axis (impact angle). Thick lines denote class~2 and 
thin lines refer to class~3. Solid lines show the sensitive area 
for particles which approach from any direction except along the 
-Z axis is; dashed lines show the sensitive area
for particles approaching along the -Z direction.
Shadowing by the magnetometer boom and cannister becomes significant for 
impact angles larger than $\rm 33^{\circ}$. 
The adopted sensitive area for class~3 particles approaching parallel to the
sensor axis is $\rm 110\,cm^2$, that for class~2 is $\rm 1,000\,cm^2$. 
}
\end{figure}

\begin{figure}
\epsfxsize=11.5cm
\epsfbox{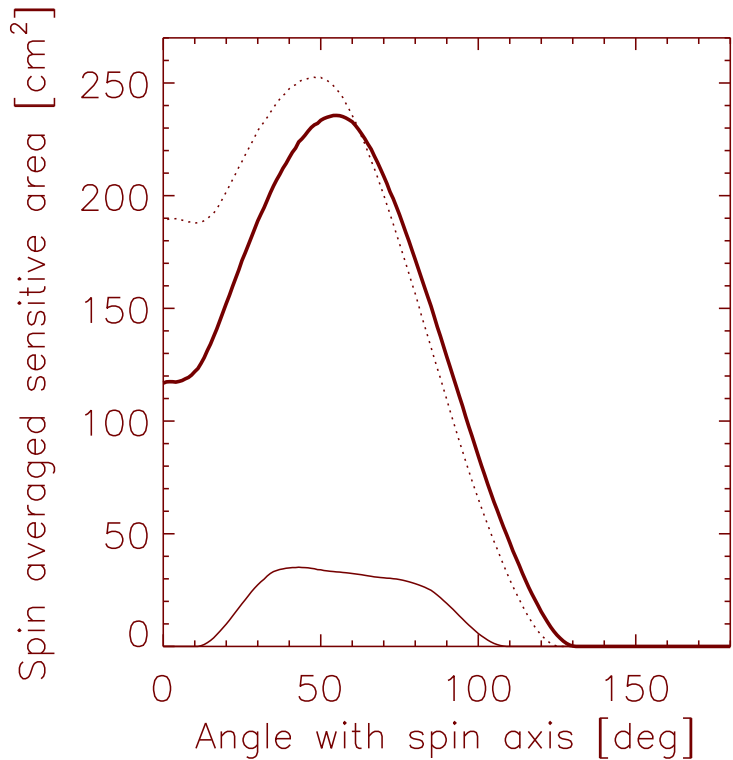}
        \caption{\label{sensarea}
Sensitive area of DDS as a function of angle with the spacecraft spin axis
averaged over one spin-revolution.
The thick solid curve  shows the total sensitive area of the sensor (class~2), the 
thin solid curve that for small class~3 impacts. 
A mounting angle of $\rm 60^{\circ}$ has been used for these curves. 
The spin-averaged sensitive area for class~2 and a $\rm 55^{\circ}$ mounting angle 
taken from Gr\"un et al. (1992b) is shown 
as a dotted line for comparison. The detector sensitivity curves shown in 
Fig.~\ref{sensareadds} (solid lines) have been adopted.
}
\end{figure}

\end{document}